# ENHANCED XML VALIDATION USING SRML

Miklós Kálmán[1] and Ferenc Havasi[1]

Department of Software Engineering, University of Szeged, Hungary

## ABSTRACT

*Data validation is becoming more and more important with the ever-growing amount of data being consumed and transmitted by systems over the Internet. It is important to ensure that the data being sent is valid as it may contain entry errors, which may be consumed by different systems causing further errors. XML has become the defacto standard for data transfer. The XML Schema Definition language (XSD) was created to help XML structural validation and provide a schema for data type restrictions, however it does not allow for more complex situations. In this article we introduce a way to provide rule based XML validation and correction through the extension and improvement of our SRML metalanguage. We also explore the option of applying it in a database as a trigger for CRUD operations allowing more granular dataset validation on an atomic level allowing for more complex dataset record validation rules.*

## KEYWORDS

*XML, XSD, XML Validation, Dataset Validation, SRML*

## 1. INTRODUCTION

Data exchange has evolved considerably over the years. Distributed systems share vast amounts of information in a matter of seconds. The most commonly used format for text based (non-binary) information exchange is XML [1]. There are many advantages to this format, however it does have its shortcomings. One of these is that since it is text based there is a possibility that the data it contains is not valid or was entered incorrectly. The structure is completely free and there is no restriction on what elements (text nodes) the user can enter. To provide a structural description the XSD [2] schema was introduced. These schemas allow the domain owners of the XML to define the structural requirements. It allows the definition of what elements the document can contain, what the attribute types are and describes the order and dependencies of elements.

Examining the exploits against sites and their databases most of them target the weakest point of these systems: data integrity and validity. Lots of the sites use XML for SOAP [3] operations or data exchange and as such validation is a very important aspect of XML data storage and transmission. Most validators can read the XSD file and validate the XML document against it. This detects most of the syntax errors, however it cannot describe more complex relationships between nodes that may be needed for validation. In an earlier article we introduced the SRML [4] language, which allowed semantic rules to be defined for attribute relationships. The metalanguage was primarily used to compact XML documents based on the rules. This opened up a plethora of possibilities in terms of describing relationships between attributes. We decided to extend this language and create an extension to the XSD format that allows these types of rules to be used during XML validation. In the process of this extension support was added for element-based rules thus simplifying the reference of nodes using the power of XPath [5]. In the earlier definition of the language referencing nodes within the context yielded unnecessary complications as it was not possible to reference all nodes and attributes. One of the most pressing issues we were faced with was how to store the rules without obstructing the XSD validation itself. The solution used was to bundle the SRML rules into the `appinfo` section of the XSD. This section is mostly used by JAXB [6] (Java XML Broker) for marshaling and unmarshalling meta information so it seemed appropriate. We have extended the standard Java XSD validator using a Spring project. The validator first runs the normal XSD validation using the XSD file provided. This validator ensures that the XML is well-formed [7]. It then reads the `appinfo` and validates the XML using the SRML rule engine. This way we get the both of both worlds. The normal XML validator will filter out the nodes/attributes, which do not conform syntactically, ensure that the XML is well-formed, and perform a type-check on the document domain. After these steps the SRML rules will validate the actual content of the nodes. This way both structure and content validation becomes possible on the XML documents.

Schematron [8] uses a similar approach to perform the validation by bundling the rules in the `appinfo` area. One of the biggest advantages our approach has over this leading validation engine is that it allows for the data to be corrected aside from just being validated. This can be very useful in environments where the validation rules can also define how to correct the input and where data loss or corruption is not an option. This allows for the input to be validated and if some items are not valid but have corresponding correction rules defined the data can be sanitized and corrected, thus allowing the data to be transmitted instead of dropping the results due to an invalid input.

We took the idea a step forward by applying the SRML validator to a database context. As most RDBMS tables and records can be represented in XML it made sense to provide a way for data validation using SRML. This approach allowed us to write the validator in a way that it can be used to validate records on insert/delete/update. The solution had its challenges, as we couldn't just apply the rules to the whole database, as that would warrant a massive memory requirement. The answer to the problem was to load parts of the records into DOM [9] trees depending on what the context of the CRUD operation was working on. This meant only parts of the records were transferred to memory and allowed the construction of a mini XML tree from the records. After the tree was built the SRML rules could be applied on this set just as if it was a standalone XML document.

The following sections will first provide some background information on the technologies as well as a brief introduction to our SRML metalanguage. We will then demonstrate the use of the new validator through an example. This example will be used in the database validation section as well to make it easier to follow. We round off the paper by showing related works, a summary and our future plans for this topic.

## 2. PRELIMINARIES

This section is dedicated to providing some color on the technologies and concepts used. We will introduce the XML format, along with the XSD schema definitions, and the SRML language. These concepts are essential to understand the later sections of this article.

### 2.1. XML

XML documents are plain text files with elements and attributes. The format is very similar to how HTML files are structured. An element can have properties called attributes; further child elements and can also contain text. Every element has to be closed off with an end tag to make it valid. A more thorough documentation on XML can be found in [1] and [10]. To demonstrate how XML documents look consider the example in *Figure 1* that stores a simple numeric expression of *3\*2.5+4*.

### 2.1.1. DTD and XSD

```
<expr>
  <multexpr op="mul" type="real">
    <expr type="int">
    <num type="int">3</num>
  </expr>
    <expr type="real">
      <addexpr op="add" type="real">
        <expr type="real">
          <num type="real">2.5</num>
        </expr>
        <expr type="int">
        <num type="int">4</num>
      </expr>
      </addexpr>
    </expr>
  </multexpr>
</expr>
```

Figure 1. XML representation of a numeric expression.

It is possible to define the syntactic structure of XML documents using DTD [11] (Document Type Definition) files and XSD (XML Schema Definition) files. DTD files can only provide the basic structure of XML files (limited to elements and attributes). Taking an XML file containing a numeric expression of *Figure 1* we can define the DTD schema in *Figure 2 (a)*.

```
<!ELEMENT num (#PCDATA) >
  <!ATTLIST num type ( real | int )#REQUIRED >
<!ELEMENT expr ( num | multextr | addexpr ) >
  <!ATTLIST expr type( real | int ) #IMPLIED >
<!ELEMENT multexpr ( expr , expr ) >
  <!ATTLIST multexpr op ( mul |div ) #REQUIRED type ( real | int ) #IMPLIED >
<!ELEMENT addexpr ( expr , expr ) >
  <!ATTLIST addexpr op ( add |sub ) #REQUIRED type ( real | int ) #IMPLIED >
```

```
<xsd:element name="expr">
  <xsd:complexType>
    <xsd:choice>
      <xsd:element name="multexpr" minOccurs="0" maxOccurs="unbounded" />
      <xsd:element name="addexpr" minOccurs="0" maxOccurs="unbounded" />
    </xsd:choice>
    <xsd:attribute name="type" use="optional">
      <xsd:simpleType>
        <xsd:restriction base="xsd:string">
          <xsd:enumeration value="int" />
          <xsd:enumeration value="real" />
        </xsd:restriction>
      </xsd:simpleType>
    </xsd:attribute>
  </xsd:complexType>
</xsd:element>
```

Figure 2. (a) DTD of numeric expression, (b) XSD snippet of numeric expression

XSD is a newer format and can do everything a DTD can, along with additional restriction definitions. The second advantage XSD files have over DTDs is that they are also XML based meaning it is easier to parse and display in a hierarchic manner. XSD documents describe the elements and their attributes just like the DTD, however also specify the type of content that the elements can have, can detail the order of elements can appear or provide a choice of elements for a given context (*Figure 2(b)*). The XSD schema can define the format of the nodes or attributes using regular expressions (e.g.: ISBN numbers or an IP address). We will detail the XSD format in more detail when we present how we extend its functionality.

### 2.1.2. Parsing XML documents

In order to perform any operations on XML files whether it is processing or validation they have to be parsed first. There are two ways of parsing an XML document: DOM (in-memory tree based) and SAX (sequential). A DOM tree is an in-memory tree that represents the whole XML document in a hierarchic manner. It allows easy parsing, query and updating of the document. Using DOM is very effective on documents that can fit in memory as it represents the whole document. *Figure 3* shows the DOM tree of the XML defined in *Figure 1*. SAX on the other hand is powerful when dealing with large XML documents which do not fit into the memory, however it is sequential therefore reading and accessing the file in one pass is not possible, neither is accessing a node directly without reading through the whole file first.

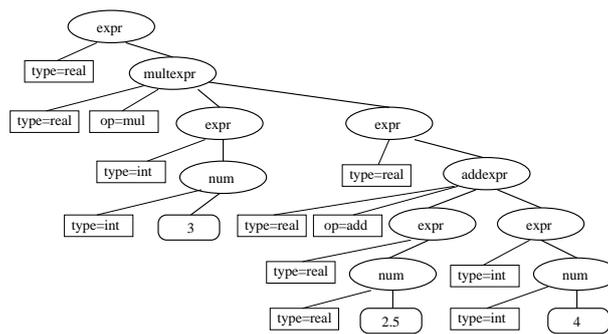

Figure 3. DOM tree of the numeric example of Figure 1.

## 2.2. SRML

We have developed a metalanguage called SRML that allows the definition of semantic rules in an XML context. Using this metalanguage we can define semantic rules that describe the relationship between XML attributes. The original definition of SRML (version 1.0) was described in [12] and [4]. We have then extended this with XPath support along with additional features. This article describes the new aspects of SRML that were introduced to enable the validation of XML documents as well as database reference descriptions. We will be using the new SRML 2.0 format throughout this article. The key differences between SRML 1.0 and 2.0 can be seen in *Figure 4*. The original definition of SRML was mostly focused on compression and the theoretical description of the rules, however nowadays the significance of compression was replaced by the importance of data validation and security.

For the validation area we decided to simplify and clean up the language to allow easier rule descriptions without sacrificing the flexibility. The new format can be used for data correction as well. The full XSD of the new format can be found in [13].

| Property | SRML 1.0 | SRML 2.0 |
|---|---|---|
| Main Focus | Compaction/Decompaction | Validation/Correction |
| Rule reference level | Attributes | Element and Attributes |
| Potential Application Area | XML Documents | XML and Databases |
| Rules based on | Attribute Grammars | AG and XPath |
| Rule Definition | Complex | Simplified with XPath |
| Numeric Expression Rules | Much overhead | Simplified, inner expression engine |
| Rule dependencies and storage | DTD and separate SRML file | Encapsulated in the XSD |

Figure 4.  Key differences between SRML 1.0 and 2.0

*Figure 5* shows how the `addexpr` section of the XML in *Figure 1* can be described in SRML 2.0. The rule definition format covers the type attribute results as well. DTDs and XSDs could not describe how the *type* attribute changes during a multiplication of an "int" and a "real". With the help of SRML 2.0 we are able to describe the type change fairly easier. Defining indexed child references is also easier, for example `../expr[1]/@type` refers to the first `expr` sibling's *type* attribute. The `../` is an extension to XPath allowing the upward navigation and reference.

The new version of SRML allows and aids the XML validation process containing several enhancements from which the following should be noted:

**XPath support:** Using XPath it is now easier to reference attributes and elements in the XML context. Previously it was a tedious job to reference specific attribute instances

**Numeric expressions:** The new format also allows numeric expression to be used during the rule context making it easier to describe expressions and use them in the rule definitions.

**Element and attribute references:** It is now possible to reference both attribute and elements. Previously SRML only operated on an attribute level

**Multiple rules for the same context:** With this new feature multiple rules can be defined for the same context. This is important for validation, as it is possible that the document can be considered valid if *"any"* of the validation rules for that context is fulfilled.

**Node relationship for tables:** SRML 2.0 introduced the option to describe database tables and thus extend the scope of the rules to the database space as well.

```
<rules-for root="addexpr">
<rule-def name="@type">
 <rule-instance>
  <expr>
   <if-expr>
    <expr>
     <binary-op op="or">
       <expr>
        <binary-op op=equal>
          <expr><value-ref path="../expr[1]/@type" /></expr>
          <expr><data>real</data></expr>
        </binary-op>
        </expr>
        <expr>
        <binary-op op=equal>
          <expr><value-ref path="../expr[2]/@type" /></expr>
          <expr><data>real</data></expr>
        </binary-op>
        </expr>
      </binary-op>
    </expr>
    <expr><data>real</data></expr>
    <expr><data>int</data></expr>
    </if-expr>
  </expr>
  </rule-instance>
 </rule-def>
</rules-for>
```

Figure 5. An SRML example for "type" attribute of the addexpr element

## 2.3. XPath

Before going into the validation one more technology has to be noted: XPath [5] (XML Path language). The XPath language is based on the DOM (tree) representation of the XML document. It provides an easy way to query for nodes and attributes using expressions. It is widely used in CSS and HTML selectors as well. This section will provide some basic information on what XPath is with a simple example. Our validation engine leverages this language heavily as it allows us to extend SRML to make element and attribute reference much easier.

The most important kind of expression in XPath is the location path. Each path is comprised of a sequence of location steps. A step element has 3 components: an axis, a node test and zero or more or predicates. The expression path is evaluated from left to right. The axis specifier describes the context of the navigation element (e.g.: child).

A node test will return all nodes in the document matching the path. Predicates allow further filtering of the results. To better demonstrate how XPath can be used consider the example in *Figure 6*. Normally the *author* attribute would be an element, but we wanted to show attribute references as well to allow better understanding of the XPath topic.

```
<books>
   <book author="J.R.R. Tolkien">
     <title>Lord of the Rings</title>
   </book>
   <book author="J.R.R. Tolkien">
      <title>The Hobbit</title>
   </book>
   <book author="Jules Verne">
     <title>Around the world in 80
days</title>
   </book>
   <book>
     <title>Anonymous</title>
   </book>
</books>
```

Figure 6.  A simple XML example for books

When using `//book/title` as the XPath query all nodes will be returned. Adding a `/text()` will only show the text content of those nodes. If we only want to query for example all books by *"Jules Verne"* we would add the `[@author="Jules Verne"]` predicate. If the predicate is supplied with a value then the expression will filter all nodes matching the given attribute. If only the attribute is specified then all nodes matching the expression where that attribute is defined will be returned. For example `//book[@author="Jules Verne"]/title/text()` will return *"Around the World in 80 days"*.

It is also possible to query nodes that have the given attribute regardless of the value. For example `//book[@author]/title` will return all titles of books which have the author attribute defined. XPath can also contain regular expressions and has a lot of in-built functions. We will not detail these, as we do not use them for our purposes.

## 3. VALIDATING XML DOCUMENTS

XML validation plays a very important role in the document's life. In many cases it is vital to ensure that an XML document is both syntactically and semantically correct. As with many text-based formats errors can arise from invalid documents. A document has to be both well-formed and valid to pass validation. The term well-formed [7] refers to the fact that all tags/elements have matching pairs, there are no overlapping elements nor do the elements/tags contain invalid characters. Once a document is well-formed the contents can be. XSD allows several ways of defining which parts of the document have restrictions and what the document has to conform to. We will use an example to demonstrate how an XSD would look like for a bookstore example. This example will then be used throughout the remainder sections to allow a better comparison.

Consider the following use case: we have a bookstore that sells books using a shopping cart. Each item in the shopping cart is a `book`, which has an *author* attribute, a `title`, an `ISBN` number, a `price` and a *cover* attribute that can be either *digital* or *hardcover*. The item also contains the `quantity` of the books in the cart and the `subtotal` for the given entry as well a `discount`. This is a simplified example as normally one would define an item with a book reference and store the quantity on that level, however to save space and avoid complexity in the example we merged these two elements into a single one. The `ISBN` number has to be a specific format and `price` can only be a number. The XML of the example can be seen in *Figure 7*.

```
<cart xmlns:xsi="http://www.w3.org/2001/XMLSchema-instance"
      xsi:noNamespaceSchemaLocation="cart.xsd" hasDiscount="false">
    <book cover="hardcover">
        <author>J.R.R. Tolkien</author>
        <title>Lord of the Rings</title>
        <isbn>1-12345-123-1</isbn>
        <qty>5</qty>
        <price>100</price>
        <discount>0</discount>
        <tax>25</tax>
        <total>625</total>
        <region>0</region>
    </book>
    <book cover="digital">
        <author>William Shakespeare</author>
        <title>Macbeth</title>
        <isbn>1-12-654321-1</isbn>
        <qty>1</qty>
        <price>100</price>
        <discount>10</discount>
        <tax>35</tax>
        <total>121.5</total>
        <region>1</region>
    </book>
</cart>
```

Figure 7.  XML of the cart example

```
<xsd:schema xmlns:xsd="http://www.w3.org/2001/XMLSchema">
<xsd:element name="cart">
    <xsd:complexType>
        <xsd:sequence>
            <xsd:element name="book" minOccurs="0" maxOccurs="unbounded" />
        </xsd:sequence>
        <xsd:attribute name="hasDiscount" type="xsd:boolean" use="optional"/>
    </xsd:complexType>
</xsd:element>
<xsd:simpleType name="ISBN-type">
    <xsd:restriction base="xsd:string">
        <xsd:pattern
        value="\d{1}-\d{5}-\d{3}-\d{1}|\d{1}-\d{3}-\d{5}-\d{1}|\d{1}-\d{2}-\d{6}-\d{1}" />
    </xsd:restriction>
</xsd:simpleType>
<xsd:element name="book">
    <xsd:complexType>
        <xsd:sequence>
            <xsd:element name="author" type="xsd:string" />
            <xsd:element name="title" type="xsd:string" />
            <xsd:element name="isbn" type="ISBN-type" />
            <xsd:element name="qty" type="xsd:integer" />
            <xsd:element name="price" type="xsd:integer" />
            <xsd:element name="discount" type="xsd:integer" />
            <xsd:element name="tax" type="xsd:integer" />
            <xsd:element name="total" type="xsd:float" />
            <xsd:element name="region" type="xsd:integer" />
        </xsd:sequence>
        <xsd:attribute name="cover">
            <xsd:simpleType>
                <xsd:restriction base="xsd:string">
                    <xsd:enumeration value="paperback" />
                    <xsd:enumeration value="hardcover" />
                    <xsd:enumeration value="digital" />
                </xsd:restriction>
            </xsd:simpleType>
        </xsd:attribute>
    </xsd:complexType>
</xsd:element>
</xsd:schema>
```

Figure 8.  XSD of the cart example

In order to ensure that all documents that get entered into our shopping cart system are valid we have to define an XSD schema for this domain. The XSD of the example can be found in *Figure 8*. We will describe each major section of this file now to show why the sections are defined as is. The XSD schema defines how the structure of the `cart` XML files needs to look like. It needs to contain a root (`cart`) element. This element has an attribute called *hasDiscount* and contains *book* child items. The `book` element definition details the elements that a `book` can have, along with their types and requires an attribute called *cover*. This attribute can take on two values: *"digital"* and *"hardcover"*. The `book` has an `ISBN` number whose format is defined with a regular expression. This ensures that all text entered into the ISBN node will need to be in the same format. Once we have the XSD document setup properly we can run a document validator on it.

During our development we used Java, as it is platform independent and a powerful language. In Java one of the ways of validating against an XSD is achieved by using the Java XML validation API. This validation will filter out invalid results and ensure that all elements are in their proper position and the types of the fields are correct. However there is no way to describe more complex validation rules. Suppose there are additional rules that need to be satisfied in order for a `cart` to be valid. For example: the `tax` on digital books should always be 0 or if the number of items in the cart is more that two then the *hasDiscount* attribute has to be true. The current XSD format does not provide a way to describe or validate against these types of conditions. This is where the power of SRML comes in. In the next section we will show how we can extend the XSD format to allow more complex validation rules.

## 3.1. Extending XSD

When trying to extend a format that is widely used one has to be careful not to break the legacy systems that are dependent on it. We had to figure out a way to stay compliant to the original XSD schema, but also allow the description and processing of SRML based validation rules. To overcome this challenge we opted to use the `appinfo` meta section of the XSD document. This section is usually used for application specific meta information storage. An example for this would be JAXB (*Figure 9*) marshalling meta overrides. JAXB [6] is Java's XML Broker which is an API used to marshall classes to and from XML. This section seemed like a viable part of the document to insert the SRML rules. The bookstore cart example described earlier will be used in the further sections along with some additional validation requirements.

```
<xsd:annotation>
    <xsd:appinfo>
        <jaxb:globalBindings collectionType="java.util.Vector"/>
        <jaxb:schemaBindings>
            <jaxb:package name="com.flutebank.custompackage"/>
        </jaxb:schemaBindings>
    </xsd:appinfo>
</xsd:annotation>
```

Figure 9. Using appinfo for JAXB binding information

## 3.2. A validation example using SRML

This section will demonstrate validation scenario of the shopping cart example mentioned earlier. **Validation Requirement #1:** the cart's *hasDiscount* attribute is true if there are more than 2 books in the cart
**Validation Requirement #2:** All books by "J.R.R. Tolkien" should receive a 20% discount
**Validation Requirement #3:** All digital books should be tax-free
**Validation Requirement #4:** The *total* entry of the book is calculated by multiplying the quantity, price and discount values

The above validation conditions would not be possible with the standard XSD format. We will now demonstrate the rules that allow the description of the validation requirements. To embed rules into the XSD first we have to define the annotation section and the `appinfo` section the following way:

```
<xsd:annotation>
<xsd:appinfo xmlns:srml=http://www.sed.inf.u-szeged.hu/SRMLSchema
    srml:schemaLocation="srml.xsd">
```

Based on the SRML XSD the top-level definition element is the `srml-def` node. This element contains one or more `rules-for` elements. The `rules-for` elements define the context root of

the rule:

```
<srml:srml-def>
<srml:rules-for root="cart" >
```

Each `rules-for` element has one or more `rule-def` entry. This specifies the target attribute or element that is to be validated. For example the following will target the cart's *hasDiscount* attribute:

```
<srml:rule-def name="@hasDiscount" mode="correct" match="any" >
```

The *mode* attribute in the above example tells the validator what to do with the results. The possible values are *"validate"* and *"correct"*. The first mode will perform the validation based on the rule and report any failures it encounters. The second mode (*"correct"*) will perform the validation and if it fails then attempt to alter the value based on the expected value defined in the rule. This is an important option, as it will ensure that the data is correct even when the validation fails. In many cases this mode can recover the XML document and allow it to be valid again. In our example the *hasDiscount* attribute value is automatically corrected if its validation fails.

The *"match"* attribute informs the validation engine what to do with multiple `rule-def` elements. If the *match* attribute is marked as *"any"* that means that the given validation rule returns true if any of the `srml-instance` rules are matched. Each `rule-def` can contain one or more `rule-instance` elements. These elements define different rules for the same context. This can be useful when a node is considered valid when any of the listed rules return successfully.

Every `rule-instance` has a `validation-error` tag and an `expr` tag. The `validation-error` element is used to pass in the string that is used when the rules in the instance fail. This string is returned to the user as a validation error which is more descriptive than throwing a validation exception. The `expr` tag contains the rule itself. *Figure 10* shows the logic behind the validation procedure.

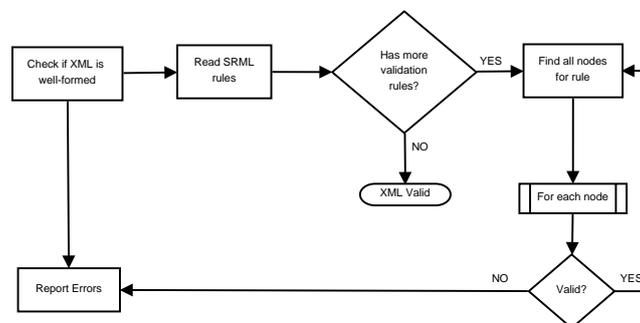

Figure 10.  XML validation process using SRML

Taking the cart example earlier we will present the logic behind the rules for each major validation requirement we mentioned along with their rule snippets.

### 3.2.1. Validation requirement #1

*Requirement: The cart's* hasDiscount *attribute is true if there are more than 2 books in the cart.*
The SRML rule for this can be seen in *Figure 11*. The rule for this is pretty straightforward; we define a root of `cart` and a `rule-def` name of *@hasDiscount*. The @ sign denotes that we will be referencing a variable for this validation rule. As all paths are converted to XPath this will reference all cart.hasDiscount attributes. The expression for the validation is a `binary-op` operation with a `greater` comparator bundled with a `count-children` element that will return the number of children with the name "book". This means the number of `book` items in the *cart*. We use using the *correct* as the primary validation mode, meaning even if the cart's *hasDiscount* attribute is incorrect the system will attempt to correct it using the expected value based on the rule definition. We can describe the validation rule in the following pseudo-form: *Validate the hasDiscount attribute of cart: Count the number of children elements with "book" as their name. If this is greater than two then return true as the validation result otherwise false. If the attribute is invalid then correct it based on the rule description.*

```
<srml:rules-for root="cart" >
  <srml:rule-def name="@hasDiscount" mode="correct" match="any" >
    <srml:rule-instance>
      <srml:validation-error>Discount value incorrectly set for cart
      </srml:validation-error>
      <srml:expr>
        <srml:if-expr>
          <srml:expr>
            <srml:binary-op op="greater">
              <srml:expr>
                <srml:count-children name="book" />
              </srml:expr>
              <srml:expr>
                <srml:data>2</srml:data>
              </srml:expr>
            </srml:binary-op>
          </srml:expr>
          <srml:expr>
            <srml:data>true</srml:data>
          </srml:expr>
          <srml:expr>
            <srml:data>false</srml:data>
          </srml:expr>
        </srml:if-expr>
      </srml:expr>
    </srml:rule-instance>
  </srml:rule-def>
</srml:rules-for>
```

Figure 11.  SRML for Validation Requirement #1

```
<srml:rules-for root="book">
  <srml:rule-def name="discount" mode="validate" match="any">
    <srml:rule-instance>
      <srml:validation-error>This book is by J.R.R. Tolkien and does not
        have the discount set to 20 percent</srml:validation-error>
      <srml:expr>
        <srml:if-expr>
          <srml:expr>
            <srml:binary-op op="equal">
              <srml:expr>
                <srml:value-ref path="../author" />
              </srml:expr>
              <srml:expr>
                <srml:data>J.R.R. Tolkien</srml:data>
              </srml:expr>
            </srml:binary-op>
          </srml:expr>
          <srml:expr>
            <srml:data>20</srml:data>
          </srml:expr>
          <srml:expr>
            <srml:instance-value />
          </srml:expr>
        </srml:if-expr>
      </srml:expr>
    </srml:rule-instance>
  </srml:rule-def>
</srml:rules-for>
```

Figure 12.  SRML for Validation Requirement #2

## 3.2.2. Validation requirement #2

*Requirement: All books by "J.R.R. Tolkien" should receive a 20% discount.* This validation requirement needs to reference the *discount* child node of all *book* elements. Since the target of the rule is not an attribute the @ sign is left out of the reference. The SRML rule for this validation requirement can be seen in *Figure 12*.

In this example all `discount` elements are validated that are beneath the `book` elements. The `binary-op` operation is used with an *equal* comparator. The `value-ref` element refers to a value returned by the expression in the *path* attribute (`../author`). This means that it is a sibling (named *author*) of the current element. The `../` path identifier will go up one level and take the element named in the second part of the path. As our current context is `book/discount` the value compare needs to take `book/author`. This is not a standard XPath identifier, but we decided to implement this to allow easier reference. The pseudo form of the rule is as follows: *Validate the book/discount element content. If the content of the author sibling element is equal to "J.R.R Tolkien" then the discount has to be 20%. If the author is different simply use the discount written in the document.*

In this example the `instance-value` element provides values for the else branches of the conditional nodes. This is important since if the validation condition does not match the actual value should be returned for the attribute/element value in question. The example basically says: if the author is *"J.R.R Tolkien"* then return 20% otherwise use the value that is present in the element/attribute that is being validated. This means that only those `book` nodes are validated whose author is *"J.R.R Tolkien"* all other books are ignored for this rule (as they will return true for this rule-instance).

### 3.2.3. Validation requirement #3

*Requirement: All digital books should be tax-free.* This validation rule will reference the `tax` element of the `book`. The condition is that if the *cover* attribute is *digital* then the `tax` value has to be *0*, otherwise the actual value will be used. The rule snippet can be seen in *Figure 13*. The figure also shows how the `value-ref` element references an attribute value. The example's path of `../@cover` refers to the parent's *cover* attribute. Since the mode is set to *"correct"* the validation rule will replace the attribute value for the `tax` to *0* if *cover* attribute of the `book` is *digital*.

```
<srml:rules-for root="book">
   <srml:rule-def name="tax" mode="correct" match="any">
      <srml:rule-instance>
         <srml:validation-error>The tax value is not correct as digital books are tax free!
         </srml:validation-error>
         <srml:expr>
            <srml:if-expr>
               <srml:expr>
                  <srml:binary-op op="equal">
                     <srml:expr>
                        <srml:value-ref path="../@cover" />
                     </srml:expr>
                     <srml:expr>
                        <srml:data>digital</srml:data>
                     </srml:expr>
                  </srml:binary-op>
               </srml:expr>
               <srml:expr>
                  <srml:data>0</srml:data>
               </srml:expr>
               <srml:expr>
                  <srml:instance-value />
               </srml:expr>
            </srml:if-expr>
         </srml:expr>
      </srml:rule-instance>
   </srml:rule-def>
</srml:rules-for>
```

Figure 13. SRML for Validation Requirement #3

### 3.2.4. Validation requirement #4

*Requirement: The total entry of the book is calculated by multiplying the quantity, price and discount values.* The final validation rule defines the `total` value of the `book`. SRML was

Figure 14.  SRML for Validation Requirement #4

```
<srml:rules-for root="book">
 <srml:rule-def name="total" mode="validate" match="all">
  <srml:rule-instance>
   <srml:validation-error>The total value is not correct!</srml:validation-error>
    <srml:expr>
     <srml:reg-eval>
       #{../qty}*#{../price}*(1-#{../discount}/100)*(1+#{../tax}/100)
     </srml:reg-eval>
    </srml:expr>
   </srml:rule-instance>
  </srml:rule-def>
</srml:rules-for>
```

extended with a regular expression evaluator engine that allows a more precise and less verbose description of this type of rule. In the earlier version of the SRML language a calculation like the above would have taken several lines of `if-expr` and `binary-op` elements and would have seriously degraded the readability. By extending SRML with the `reg-eval` element it is now possible to define mathematical expressions much easier than before. The snippet for his requirement can be seen in *Figure 14*. The validation rule for `total` uses `#{..}` markers. These inform the expression engine to evaluate the node value with the path expression inside. It is also possible to reference attributes by using the @ marker in the *path* value. The `../` accessor is also available in these cases. The engine looks up the node/attribute values and replaces them in the expression after which it will evaluate the results.

### 3.3. Using SRML in the field of Databases for Dataset validation

Another large area where the validation engine can be leveraged is the field of databases. Nowadays database validation is almost as important as validating the data transmitted from one system to another. Normally semantic validation is done by type-forcing of table columns. This means that if you try to insert a string into an INT column then the database engine will report an error. Database tables use pre-defined schemas to ensure that they always contain all fields that are required. One may notice similarities between how RDBMS systems handle and store data to what we outlined with the XSD section of this article. Databases can use triggers to perform any input validation. A trigger is a function stored inside the database itself that will take the input parameters of the actual select/insert/update/delete operation and perform an operation on them. Triggers are usually used in creating audit trails of data modification or used to change the content being inserted. With triggers however it is very complicated to define validation type rules on what the data should be in context to the already existing records.

We decided to take an approach where we dynamically build up a context tree (mini XML) for the given record and allow SRML rules to be executed on them including XSD type restrictions. This opened up a plethora of possibilities with data modeling and validation. As we have written our SRMLXsd in Java it made sense to choose a database platform which allows the utilization of the codebase we created for the XML validator. We chose H2 [14] as it is a high performance RDBMS database written purely in Java. It has all the features of major RDBMS systems, but has the benefit of allowing Java classes to be defined as Triggers. *Figure 15* shows the validation procedure for databases. The contents in this section are applicable to all RDBMS systems that allow code to be executed as triggers (e.g.: H2, Oracle, Sybase, Microsoft SQL Server).

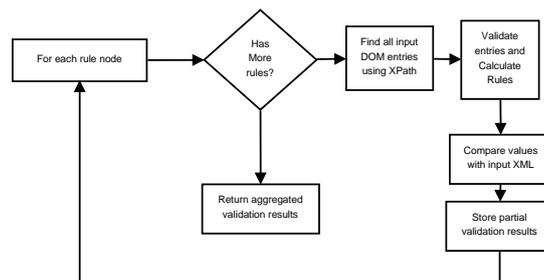

Figure 15. Validating database records using SRML

There were several challenges we faced during the implementation. The first one was to extend the rule schema to describe how the tables are linked together hierarchically. The reasoning behind this was that most database tables have references to other tables and columns (foreign key relationships). To accommodate this we introduced the `database`, `tables`, `table`, `references`, `reference` tags. With this extension a multi-tier validation approach is now possible in the SRML space (similar to *Figure 16*). We will describe the extensions first in short before proceeding further in order to provide a better understanding how these can be used to convert a flat database record into a DOM like tree.

**database** : The database section is used to store all database related relationship and definition information. It contains a `tables` and `references` element.

**tables** : This element is used to describe the keys of the tables. These keys are used to identify the rows during the creation of the DOM tree.

**table** : The table defines the keys of the tables with a *name* and *key* attribute. These are used to identify the nodes and map them to tables.

**references** : This element contains one or more reference element. It is used to store the reference relationships between tables.

**reference** : The reference element has *root*, *root key*, *child* and *child key* attributes. These values are used to map out the relationship between records of multiple tables. The keys defined in the table section are used to match the *child key* entries. This is basically a foreign key mapping resolution.

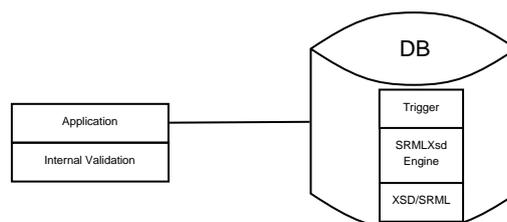

Figure 16. Multi-tier validation for applications

Using the above elements it is possible to build up a DOM tree context for a given row. This is a very powerful addition as this allows reference to other tables and their columns using SRML rules. Database related SRML rules have one restriction: as tables have columns and no hierarchic datasets all `rule-def` references need to use attribute contexts (attrname). For example the *book/author* path maps to a book table's author column. Using the previous bookstore example we will demonstrate how SRML rules can be defined as triggers on CRUD operations.

### 3.3.1. Defining table relationships

In the cart example used previously we had 2 elements: `cart` and `book`. Each element had several children and attributes. For the database example these elements were flattened into two tables as visible in *Figure 17*.

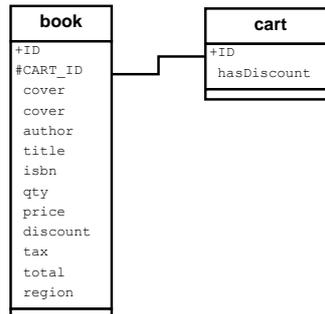

Figure 17. Database schema of cart

The SRML definition of the table relationship for the cart example can be seen in *Figure 18*. This defines the primary key of `cart` and `book` and defines a foreign key relationship between *book.CART_ID* and *cart.ID*. First we have to define the tables that would take part in context along with their primary keys. This is done with the `srml:table` element. In our example we have a `cart` and `book` table, both with *ID* as their primary keys. The next section of the definition is the `srml:references` element. This defines the foreign key relationship between the `cart` and `book` tables. The *root* is the referred table containing the *KEY* that creates the relation, allowing a DOM tree to be built from the resulting dataset. The generic query that builds this DOM tree would look like: *SELECT * FROM book WHERE book.CART ID=cart.ID*. The resulting columns are loaded as attributes along with their values into the DOM tree. This is an important thing to mention, as only the required context will be loaded during the validation. As the CRUD operation is affecting a single row at a time (even if it is part of a transaction), the context would only load the required records into the XML DOM tree.

```
<srml:database>
  <srml:tables>
      <srml:table name="cart" key="ID" />
      <srml:table name="book" key="ID" />
  </srml:tables>
  <srml:references>
  <srml:reference root="cart" root_key="ID" child="book"
      child_key="CART_ID" />
  </srml:references>
</srml:database>
```

Figure 18. Table relationship using SRML

### 3.3.2. Setup Trigger and store in database

We defined a class that can be used as the trigger for all *update*, *select*, *insert*, *delete* operations which is able to leverage the SRML rule engine for data validation. This class has an overridden function that gets called with the old and new rows that the CRUD operation pretends to. The trigger's classes along with all related classes are put on the database's classpath so that it can be accessible during use.

### 3.3.3. Store SRML XSD inside database

In order for the rules to be accessible by the triggers they need to be stored in a local table in the database. To achieve this the XSD file is persisted into a table. This XSD not only contains the SRML rules but also any other XSD restriction we may want to place on the operations. This is useful as for example we can define what values a given column can take using standard XSD restrictions and use the engine to validate them as well during the row validation. This opens up many possibilities, as normal RDBMS systems do not have a concept of enums for example.

### 3.3.4. Perform row validation using the engine

When all the pre-requisites are in place the validation is achieved automatically with the trigger hook. During CRUD operations (we can define exactly what type of operations the trigger should fire on) the database system will invoke the trigger class and pass in the *previous* row and the new row from the operation. The *previous* row is passed in when an update is being done or when a delete occurs. When the system is performing and insert, then the *previous* row is null. Based on the rows we look up any affecting validation rules from the SRML set and construct the DOM tree using the reference elements. This DOM tree is assembled using multiple select operations with the reference elements defined (foreign keys). The resultset is then converted into a DOM tree where the attributes of each element are the columns of the table and the nodes themselves are the rows. Taking the rows from the tables in *Figure 17* the system will build a DOM tree described in *Figure 19*.

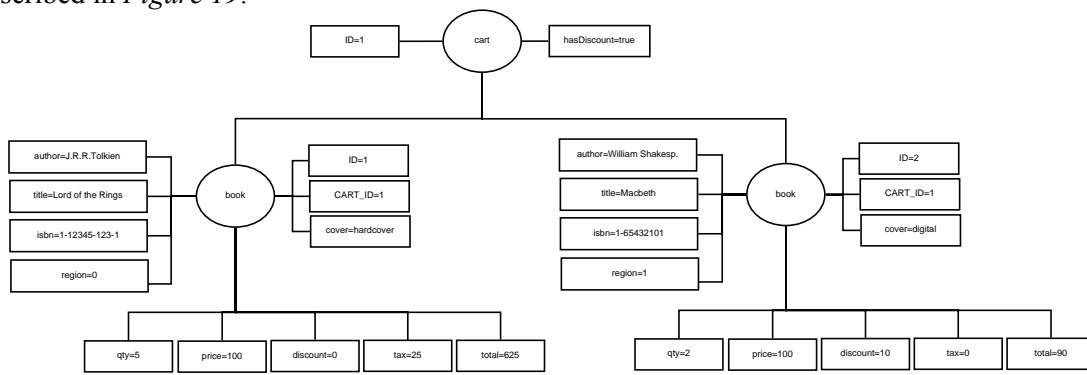

Figure 17. DOM tree of the cart database schema

After the DOM tree is constructed from the resultset context the validation proceeds as previously described. If the resultset is not valid an exception is thrown with the text of the `validation-error` element. This allows the user of the database system to see what the validation error was, for example like the one in *Figure 20*. If the rules had a mode of *correct* then the values are corrected prior to reporting an error. This allows High Availability and Data Oriented systems to retain as much data as possible by correcting input. Data corruption can also happen during network related transmission making this approach a viable candidate for validation in those fields as well.

```
Error: Validation Error. Message=[The total value is not correct!].  Found=[1625.0]. Expecting=[1125.0];
SQL statement: insert into BOOK (CART_ID,COVER,AUTHOR,TITLE,ISBN,QTY,PRICE,DISCOUNT,TAX,TOTAL)
VALUES(1,'hardcover','J.R.R. Tolkien','Lord of the Rings','1-12345-123-1',5,100.0,0,125.0,1625.0) [0-169]
SQLState:  null ErrorCode: 0
```

Figure 20. Cart database schema validation exception

## 4. RELATED WORK

XML validation has always been a hot topic amongst the community. There are several advances in this field. Most validators however only concentrate on semantic validation and do not offer rule based validation scenarios. Currently there are two major pattern/rule based validation projects available that resemble our SRML based approach. Most of the approaches are very well defined and we could have taken one of them as the basis for our extension. The main reason behind going with our own format was that we have defined the language previously and it has a potential to become a complete solution for both XML validation and correction.

The first project to mention is the RelaxNG [15] project. It can be considered as the one of the earliest of schema validators. It has a compact syntax and the document is well-defined. It contains non-deterministic content models, however it does not provide any datatype support and has no support for the XSD numeric occurrence constraints (in XSD it is possible to specify the *minOccurs/maxOccurs* attribute which will inform the validator of the quantitative property). In RelaxNG the attributes are defined as part of the content model providing a homogeneous view of the XML tree, similarly to how the DOM tree represents the XML tree. RelaxNG was a merge between Relax and TREX (Tree Regular Expressions for XML). *Figure 21* shows a simple rule definition of a simplified book-cart example in RelaxNG format. The definition is similar to how XSD defines the structure however does not offer data correction out of the box making the SRML a better option for this purpose.

```
<start>
    <element name="cart">
       <zeroOrMore>
          <ref name="book_entity"/>
       </zeroOrMore>
    </element>
</start>
<define name="book_entity">
    <element name="book">
       <attribute name="cover" > </attribute>
       ...
    </element>
</define>
```

Figure 21. RelaxNG example

One of the most known pattern based validators available is the Schematron [8] project which was also recorded under ISO/IEC 19757-3:2006. The authors of this project initially started out by extending the Word UML format used by Microsoft products [16][17] and introduced a language to model the relationships. This approach allows many types of structures to be represented and allows the developer to perform reporting and assertions on these. The project can also use XPath for finding nodes. This approach is focused on validating XML files using rules and assertions. It is very powerful, but lacks the option to correct the input. Our approach not only validates using rules, but also allows for the XML document to be corrected if the rule definitions allow for it. The subset of the example (total value calculation) in Schematron can be seen in *Figure 22*.

```
<pattern name="Book Total value check">
    <rule context="book">
       <assert test="total != qty * price * (1- discount /100)*(1+ tax/100)">
          Total Mismatch
       </assert>
    </rule>
</pattern>
```

Figure 22. Schematron example for value validation

Another project that should be mentioned is CAM [18], which is short for Content Assembly mechanism. CAM is different from other approaches as it does not define complex grammars, but rather approaches the validation from a structural pattern-matching front. The language allows business rules to be defined using XPath references and corresponding actions (e.g.: *condition* : *string − length*(.) < 11 *action* : *setDateMask*(*YYYY − MM − DD*) ). It also allows cross- and current-node conditional validation (e.g.: quantity needs to be between 1 and 100). The rule definition is more compact than SRML, however it still lacks the data correction feature our approach allows for.

## 5. SUMMARY AND FUTURE WORK

We started out by introducing the background knowledge needed to understand article seasoned with references to how we will be leveraging them in our solution. We have shown how we used SRML 2.0 to extend the XSD format to allow for both syntactical and content validation. The aspects of the new SRML format can be summarized the following way:

1. Allows both attribute and element references.
2. Integrates into the appinfo section of the XSD making it easier to deploy
3. The new format focuses on XML validation in contrast to its predecessor which focused on making the XML documents smaller
4. The new validator engine leverages the Java XML Validator, which ensures the well-formedness of the input files aside from the additional validation rules that can be defined on the context of the content itself.
5. Leverages XPath to reference nodes and their values
6. Allows the definition of complex validation rules, including regular expressions
7. Allows the XML document to be corrected using the new SRML rules
8. Potentially usable in an RDBMS environment. This allows the datasets to be validated using SRML prior to being inserted into the database using triggers. The datasets and their context are built up using a mini-DOM tree allowing the SRML rules to be applied to them. This allows dataset references to existing rows and columns as well.

In the future we are planning to extend the capabilities of the tool further by exploring possibilities of applying validation to larger documents using SAX and potentially for BigData using Hadoop with MapReduce techniques. We will also investigate how to simplify our SRML format even further to make it use less space in the document without compromising the functionality and readability. We would also like to venture more onto the RDBMS usage and potentially create a query augmentation as well, which would allow SRML rules to be applied to result set using stored procedures. We also plan to implement an update to Schematron to allow SRML rules to be inserted into its ruleset. We will also aim to extend the CAM specification as well to make it possible to convert SRML rules to and from the CAM format making the transition between formats easier. We are also thinking of extending the language even further to other areas like form validation, advanced data correction.

**Authors**

**Miklós Kálmán** received the diploma from the University of Szeged as a Software Engineer mathematician in 2003. He is currently a PhD student at the Department of Software Engineering where his research area is XML compaction and validation. He has experience in low-level hardware programming, artificial intelligence and image processing. He is well versed in Enterprise systems and in Java development as well as Database systems.

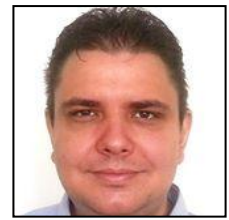

**Ferenc Havasi** received his diploma from the University of Szeged as a Software Engineer mathematician in 2001. He is currently a PhD student at the Department of Software Engineering where his research areas are XML compaction and code compression.

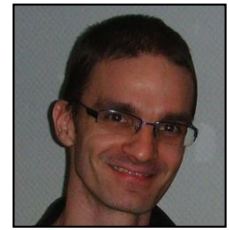